\documentclass[journal]{IEEEtran}

\usepackage{tikz}
\usepackage[utf8]{inputenc}
\usepackage{pgfplots} 
\usepackage{pgfgantt}
\usepackage{pdflscape}
\pgfplotsset{compat=newest} 
\pgfplotsset{plot coordinates/math parser=false}
\usepackage{url}
\usepackage{etoolbox}

\usepackage{standalone}

\usepackage{dblfloatfix}

\usepackage{fixltx2e}
\usepackage{amsmath,amssymb}
\usepackage{amsfonts}
\usepackage{cite}
\usepackage{tikz}

\ifCLASSINFOpdf
\else
\fi

\makeatletter
\patchcmd{\@maketitle}
  {\addvspace{0.5\baselineskip}\egroup}
  {\addvspace{-0.15\baselineskip}\egroup}
  {}
  {}
\makeatother

\begin{document}
\bstctlcite{IEEEexample:BSTcontrol}
%
\title{Applying Neural Networks in Optical Communication Systems: Possible Pitfalls}
%
%
%

\author{Tobias A. Eriksson, Henning B\"ulow, Andreas Leven
\thanks{Manuscript version 2.0. Submitted June 12, 2017}
\thanks{T. A. Eriksson, H. B\"ulow and A. Leven
       are with Nokia Bell Labs, Lorenzstr. 10, 70435 Stuttgart, Germany (email: tobias.eriksson@nokia-bell-labs.com).}
       }


\maketitle

\begin{abstract}
We investigate the risk of overestimating the performance gain when applying neural network based receivers in systems with pseudo random bit sequences or with limited memory depths, resulting in repeated short patterns. We show that with such sequences, a large artificial gain can be obtained which comes from pattern prediction rather than predicting or compensating the studied channel/phenomena. 
\end{abstract}


%
\IEEEpeerreviewmaketitle

\vspace{-3pt}
\section{Introduction}
\IEEEPARstart{A}{rtifical} neural networks (NNs) and deep learning constitute one of the hottest research topics at present \cite{lecun2015deep}. We use services that rely on deep learning daily in for instance translation services \cite{bahdanau2014neural, sutskever2014sequence}, image recognition \cite{krizhevsky2012imagenet}, face recognition \cite{taigman2014deepface}, speech recognition \cite{hinton2012deep}, etc. One of the key strength of NN based techniques over classical machine-learning techniques is that an NN can be fed with raw data and automatically find the feature representation that is needed for, say, classification \cite{lecun2015deep}. NNs can approximate complex nonlinear functions \cite{le2015tutorial} and are especially useful for problems that cannot be easily described analytically or with a model. For telecommunication systems, NNs have been proposed for various funciontions such as channel equalization \cite{burse2010channel, peng1991adaptive}. One goal, which in some sense is very close to true artificial intelligence, is to be able to learn new signal processing algorithms using for instance neural turing machines \cite{graves2014neural}.

Recently, research on machine learning techniques for optical communication systems have been increasingly popular for applications such as nonlinearity mitigation and carrier phase recovery \cite{zibar2016machine} and in the last few years, NNs have started to find their way into optical communication systems. NNs have been applied in orthogonal frequency division multiplexing (OFDM) systems to compensate for nonlinear propagation effects in single channel systems \cite{jarajreh2015artificial, chen2006channel, Giacoumidis15, ahmad2016radial}. Further, NNs have been applied in direct detection optical systems to compensate for both linear and nonlinear distortion \cite{estaran2016artificial}. For directly modulated lasers systems with direct detection, NNs have been proposed as a method to pre-compensate the signal to increase the tolerance to chromatic dispersion \cite{warm2009electronic}. Furthermore, NNs have also been proposed for equalization in 8-ary pulse amplitude modulated (8PAM) direct-detection systems \cite{gaiarin2016high}. In \cite{wang2017system}, NNs are proposed to mitigate various system impairments for coherent polarization-multiplexed (PM) 16-ary quadrature amplitude modulation (QAM). Similarly, NNs have been applied to mitigate nonlinear effects such as self-phase modulation for 16QAM \cite{owaki2016equalization} and QPSK signals \cite{shen2011fiber}. Furthermore, NNs have been used to estimate the channel probability density function in direct-detection optical systems \cite{rios2017experimental}. For indoor wireless optical communication systems, NNs have been proposed for adaptive equalization of various modulation schemes \cite{rajbhandari2010application}. Other uses in optical communication systems include modulation format recognition \cite{khan2012modulation} and optical performance monitoring \cite{wu2009applications, ribeiro2012optical}.

In this paper, we investigate the risk of overestimation when applying neural network based methods in optical communication systems. We show that when using pseudo random bit sequences or short repeated sequences, the gain from applying neural network assisted receivers can be severely overestimated due to the capability of the NNs to learn to predict the pattern that is used.

\vspace{-7pt}
\section{Systems with Neural Networks}\vspace{-1pt}
Most experimental investigations and some simulation investigations for optical communication systems rely on transmission of pseudo-random bit sequences (PRBS) such as PRBS sequences \cite{PRBS,itutPRBS} or De Bruijn sequences. For conventional studies without NNs or similar nonlinear classification or estimation methods, such types of sequences are preferred over purely random sequences. The reason for this is the memory constraints. In experiments, endlessly repeated sequences are used based on, e.g., digital-to-analog converters (DACs) or pulse pattern generators (PPGs) with limited memory depths. Using PRBS, the statistics is well approximated to a truly random sequence which would not be the case of a repeated instance of one truly random sequence with limited length. 

However, for NNs we will show that the use of PRBS sequences can lead to overestimation of the NN performance, simply because the NN is capable of learning to predict the pattern. The PRBS is generated using linear shift registers with for instance the following polynomials
\begin{equation}\vspace{-1pt}
B_{\textrm{PRBS7}} = x^7 + x^6 +1,
\end{equation}\vspace{-1pt}
and
\begin{equation}
B_{\textrm{PRBS15}} = x^{15} + x^{14} +1.
\end{equation}
The length of the PRBS sequences are $2^N-1$ where $N$ is the PRBS order. For simplicity, we want to work with power-of-two length sequences and hence we extend the length of each investigated sequence with a ``0''.

\begin{figure}[!tb]
\centering
\includegraphics[width=0.9\columnwidth]{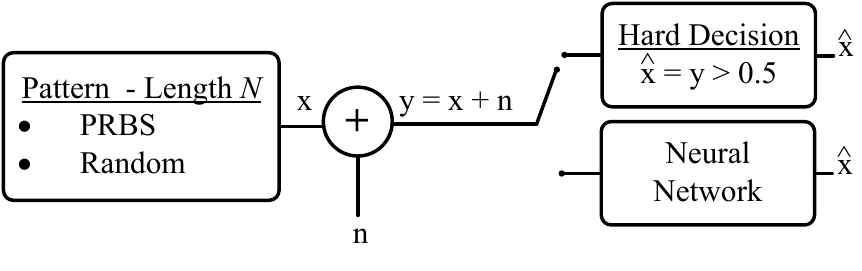}\vspace{-5pt}
\caption{Illustration of the AWGN channel with PRBS or random bits.}
\label{Fig_awgnChannel}
\end{figure}

\section{AWGN channel with repeated patterns}
\begin{figure}[!tb]
\centering
\includegraphics[width=0.9\columnwidth]{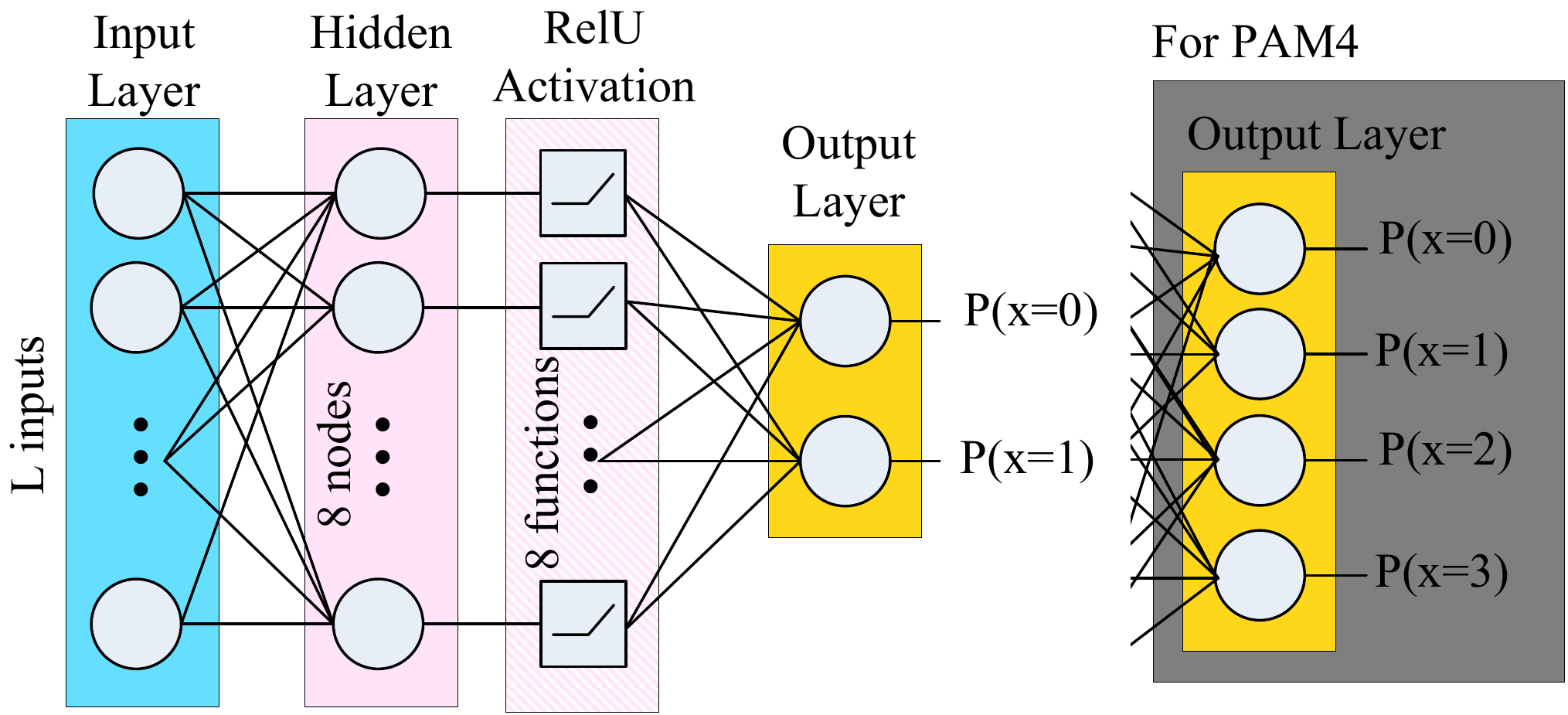}\vspace{-5pt}
\caption{Illustration of the neural network used in this paper.}\vspace{-12pt}
\label{Fig_NN_simple}
\end{figure}

To start with, we will investigate a simple scenario, i.e. binary transmission over the additive white Gaussian noise channel (AWGN) as illustrated in Fig.~\ref{Fig_awgnChannel}. We use a simple NN as illustrated in Fig.~\ref{Fig_NN_simple}. The NN consists of one hidden layer with 8 nodes and conventional rectified-linear and leaky (ReLU) activation functions. The input layer has $L$ input bits chosen symmetrical around the center bit that is estimated, i.e. $L$ is always an odd number. For the binary investigation, the output layer consists of two output nodes which corresponds to the probability of a 0 or a 1 being transmitted respectively. For the PAM4 investigation, the output layer has four nodes where the outputs corresponds to the probabilty of each transmitted PAM4 level. The network is trained using back-propagation with Nesterov’s accelerated gradient \cite{nesterov1983method}. This method is similar to stochastic gradient descent but with the gradient taken on the weights with added momentum. The loss is calculated using multinominal logistic loss. Note that we are not trying to optimize the structure of the NN, the activation functions and the training strategy, but we are rather using a simple structure for demonstration purposes.

The network is trained from scratch for different input sizes $L$ using PRBS7, PRBS15 or a repeated "random" pattern with length $2^7$. For training, noise is added with a signal-to-noise ratio (SNR) of 10 dB. At this point the BER is around $1.3\times10^{-2}$ for hard decision. The length used for the training is $2^{19}$ blocks of length $L$. For testing the neural network, we use at least $2^{16}$ input blocks of either repeated PRBS sequences or instances of a random patterns for each scenario. We always use a new realization of both the noise and the random pattern.

\begin{figure}[!tb]
\centering
\includegraphics[width=\columnwidth]{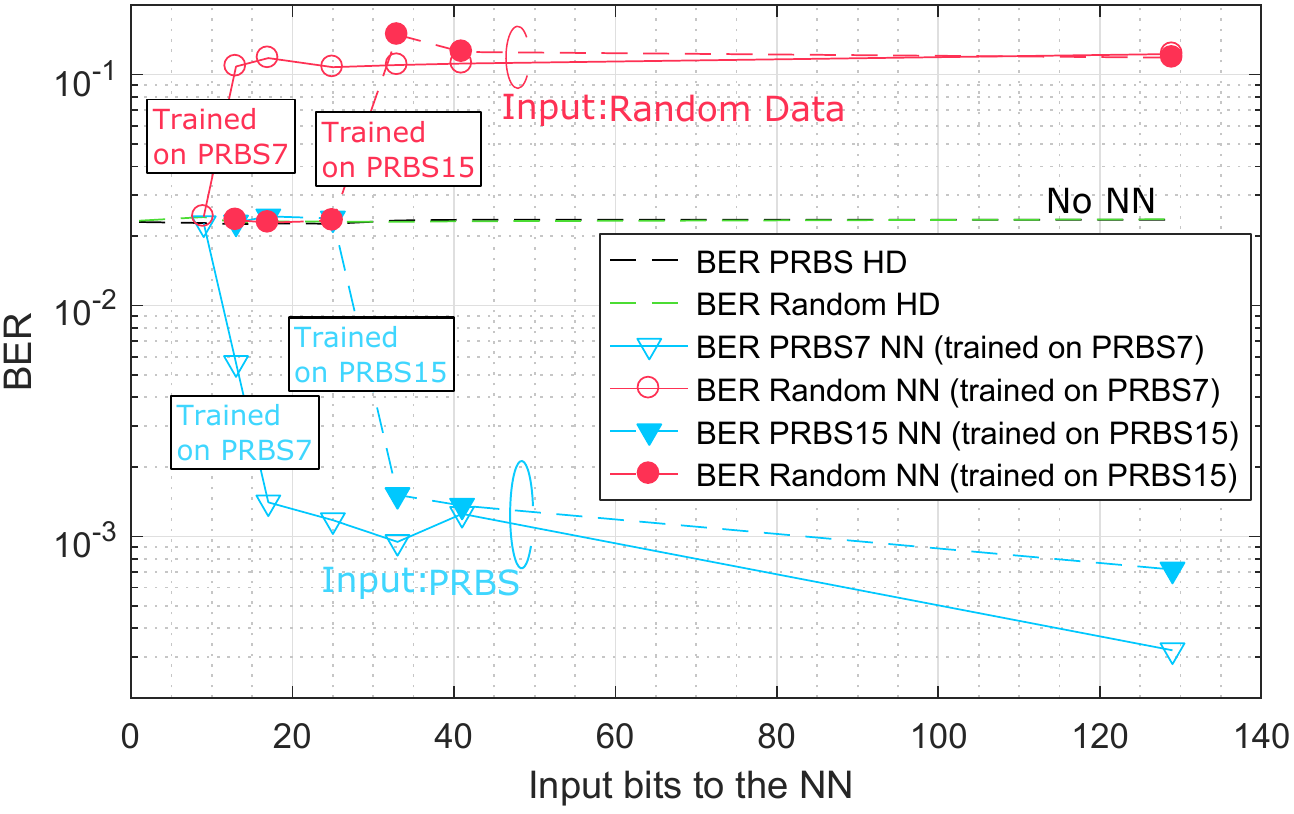}\vspace{-5pt}
\caption{BER as a function of the number of input bits to the neural network trained with either PRBS7 or PRBS15 at 9~dB SNR. Also shown is the BER for hard-decision (without NN) and the BER when random data is used as an input to the neural network trained on either PRBS7 or PRBS15.}
\label{Fig_sweepL_PRBS7_10dB}
\end{figure}

In Fig.~\ref{Fig_sweepL_PRBS7_10dB}, the BER as a function of the number of input bits to the NN is plotted at an SNR of 9~dB for the case when the network is trained with either a PRBS7 or PRBS15 pattern. The dashed lines corresponds to hard decision and is obviously independent on the block length and pattern type. When evaluating the NN with a PRBS7 sequence, we see that for 13 input bits and higher, the NN can start to predict the pattern and we see a massive decrease in the BER. For PRBS15, we need slightly longer blocks of data into the NN. For 33 input bits and above, we again start to predict the pattern. However, if we instead evaluate the NNs with realizations of a random patterns, the NNs instead increase the BER. This can be understood by the fact that the NN expects the structure of the PRBS. This shows the danger of using a PRBS pattern when evaluating NNs. I.e. the gain from the system that is being studied might be overestimated since a large portion of the gain can come from predicting the PRBS pattern that is used. 

\begin{figure}[!tb]
\centering
\includegraphics[width=\columnwidth]{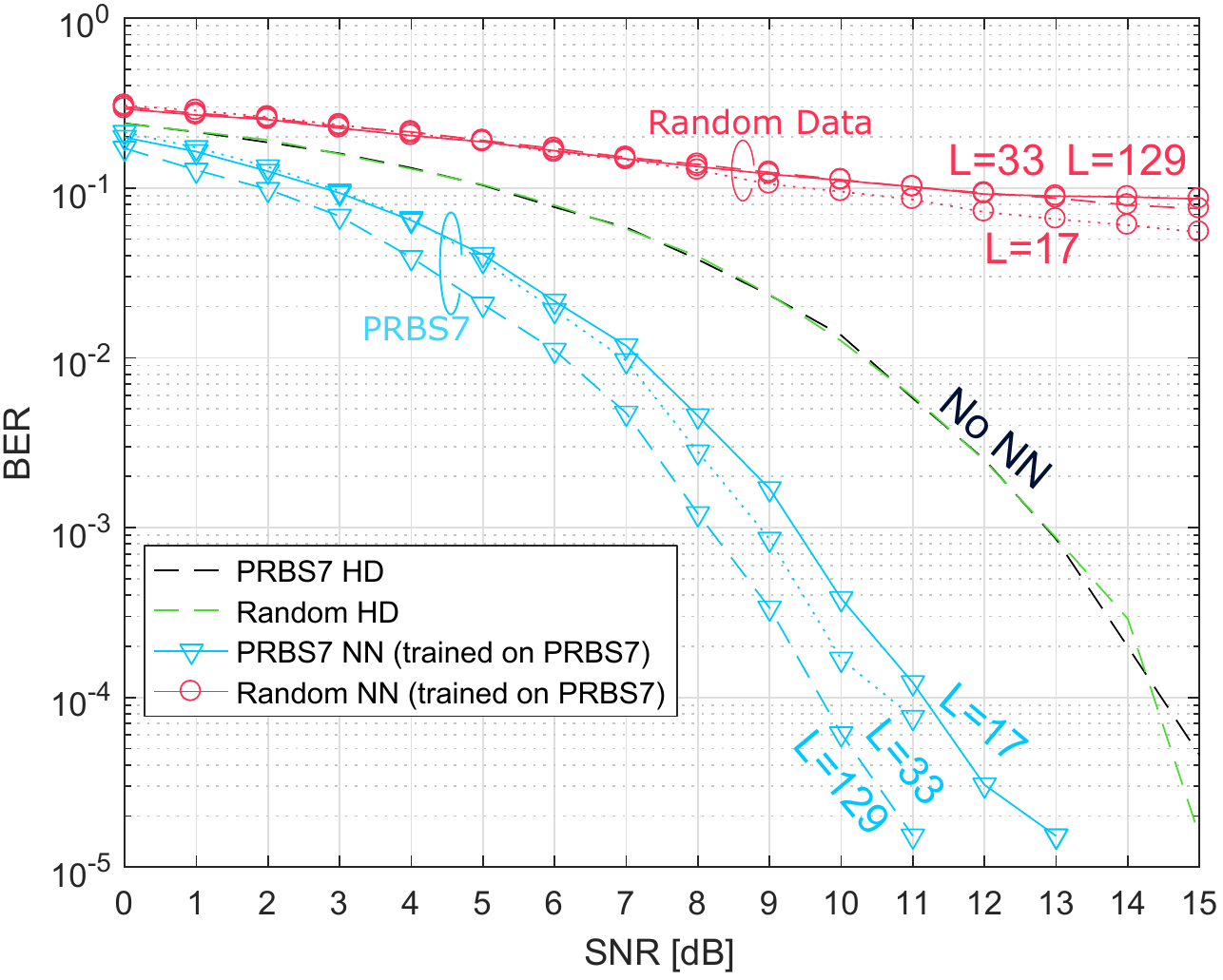}\vspace{-5pt}
\caption{BER as a function SNR for hard decision on a PRBS7 and a random pattern, and for a neural network trained on a PRBS7 with either PRBS7 or random data as input. The input length to the neural network is either 17, 33 or 129 bits symmetrical around the center estimated bit.}
\label{Fig_sweepSNR_PRBS7_L6}
\end{figure}

\begin{figure}[!tb]
\centering
\vspace{-10pt}
\includegraphics[width=\columnwidth]{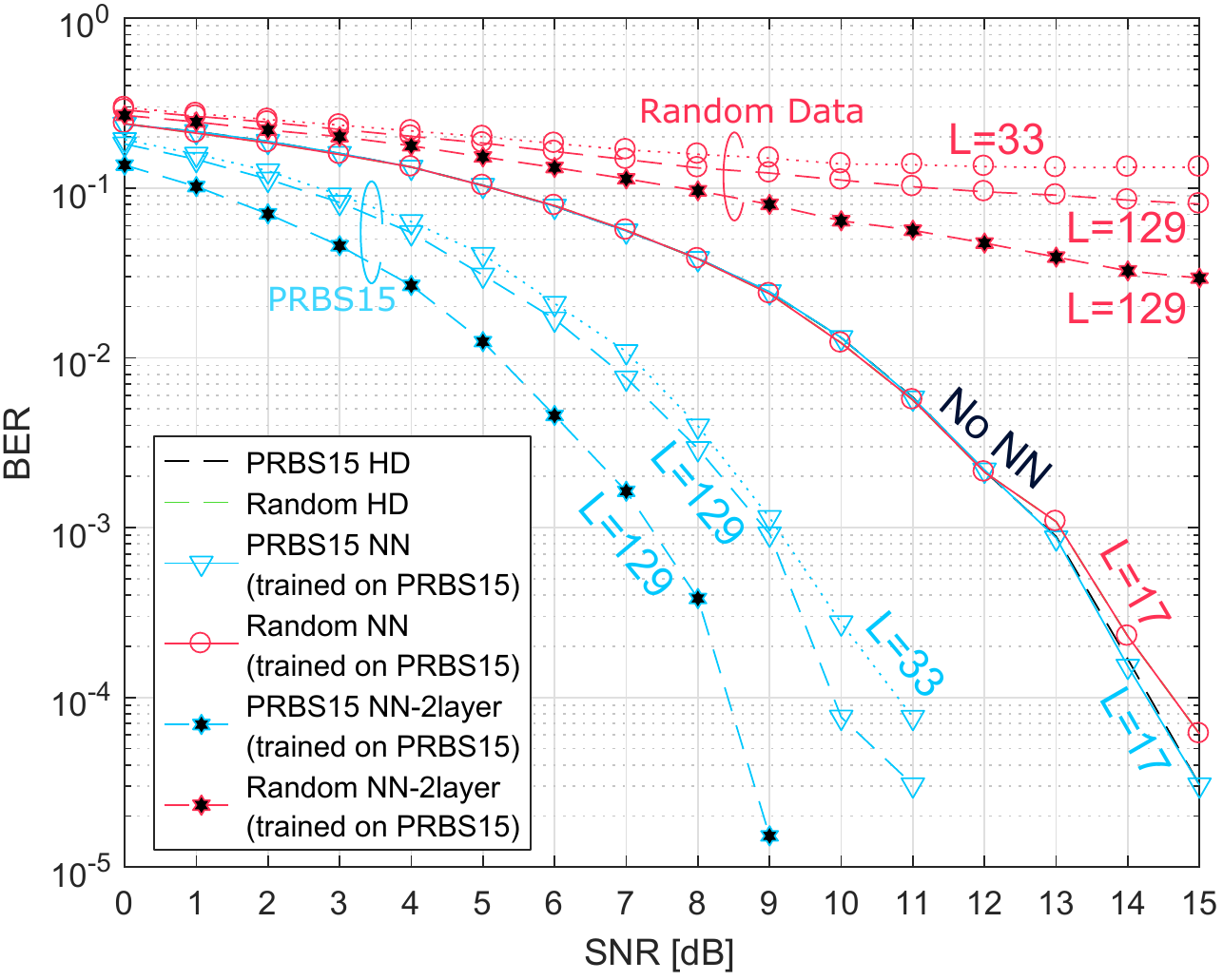}\vspace{-5pt}
\caption{BER as a function SNR for hard decision on a PRBS15 and a random pattern, and for a neural network trained on a PRBS15 with either PRBS7 or random data as input. The input length to the neural network is either 17, 33 or 129 bits symmetrical around the center estimated bit. We also plot the BER vs SNR for a second neural network which used two hidden layers with 64 nodes each. }
\vspace{-10pt}
\label{Fig_sweepSNR_PRBS15}
\end{figure}

\begin{figure}[!tb]
\centering
\includegraphics[width=\columnwidth]{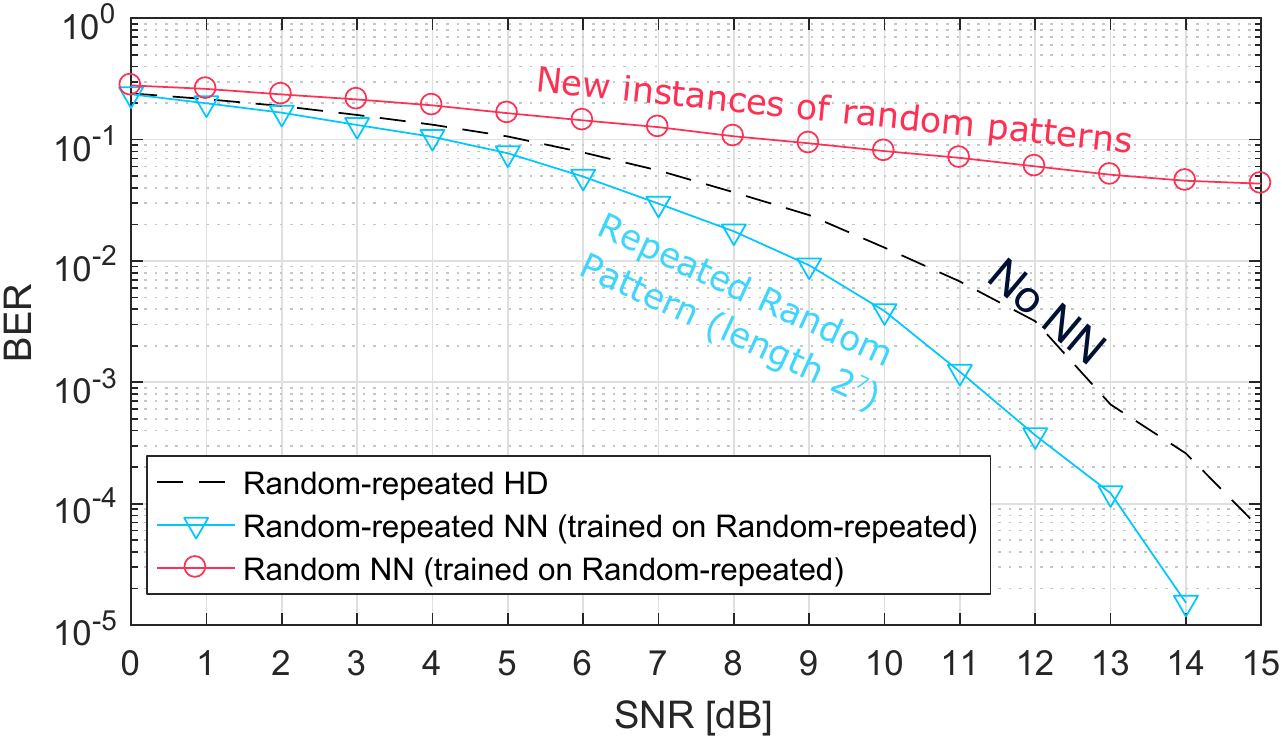}\vspace{-5pt}
\caption{BER as a function SNR for hard decision on a random  pattern, and for a neural network trained on an instance of a random repeated pattern of length $2^7$ with either the same repeated pattern or long random data sequence (length $2^{16}$) as input. The input length to the neural network is 33 bits.}\vspace{-9pt}
\label{Fig_sweepL_randomRepeated}
\end{figure}

To further illustrate this point, we fix the input length to either 13 or 25 bits and sweep the SNR for the NN trained with PRBS7 data. The results are shown in Fig.~\ref{Fig_sweepSNR_PRBS7_L6} together with the case when the NN is evaluated with random data and the results for hard decision without the NN. With this setup, we are overestimating the performance with 1.7~dB in SNR for 13 bits input and 3.3~dB for 25 bits input, both measured at BER = $10^{-3}$. Using the random pattern input gives a much worse BER for any channel SNR. 

In Fig.~\ref{Fig_sweepSNR_PRBS15}, we perform the same evaluation but for the NN that has been trained on a PRBS15 pattern. When we use a very short block length (17 in this case) into the neural network, we get identical performance as for the hard decision results. In other words, for this short block length the very simple neural network structure applied in this paper cannot predict the pattern and it finds the optimal hard decision threshold. For longer block lengths, 33 and above, the neural network again can predict parts of the pattern and the system will overestimate the performance. To show that the overestimation is dependent on the NN layout, we apply a second neural network which is using two hidden layers with 64 nodes each, and as seen this network is better at predicting the PRBS15 pattern and thus yields an even larger overestimation of the gain. 

Finally, we train the NN on a repeated instance of a random sequence with length $2^7$. This is an extremely short pattern, but we use this as an illustrative example as our neural network is also very small with, in this case 33 input bits. The prediction results are shown in Fig.~\ref{Fig_sweepL_randomRepeated}. Even in this case, we are over estimating the performance even though the pattern used is not generated with linear shift registers like in the PRBS case. We conjecture that we would also have overestimation for long repeated sequences, with more complicated NNs which longer inputs and/or more hidden layers.

\begin{figure}[!tb]
\centering
\includegraphics[width=0.93\columnwidth]{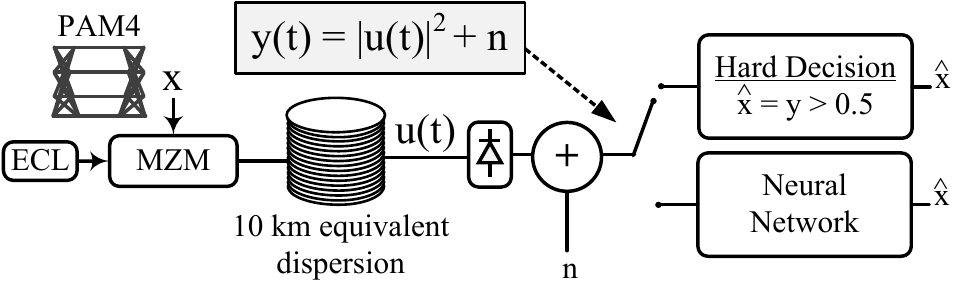}
\caption{Simulation setup used to illustrate the dangers of overestimation in a real application. We study the estimation of transmitted symbols in presence of the nonlinearity introduced by dispersion and square-law detection. We use 32~Gbaud PAM4 signals and 10~km worth of dispersion. The noise is added after the photodetector.}
\label{Fig_simulationSetup}
\end{figure}

\begin{figure}[!tb]
\centering
\includegraphics[width=\columnwidth]{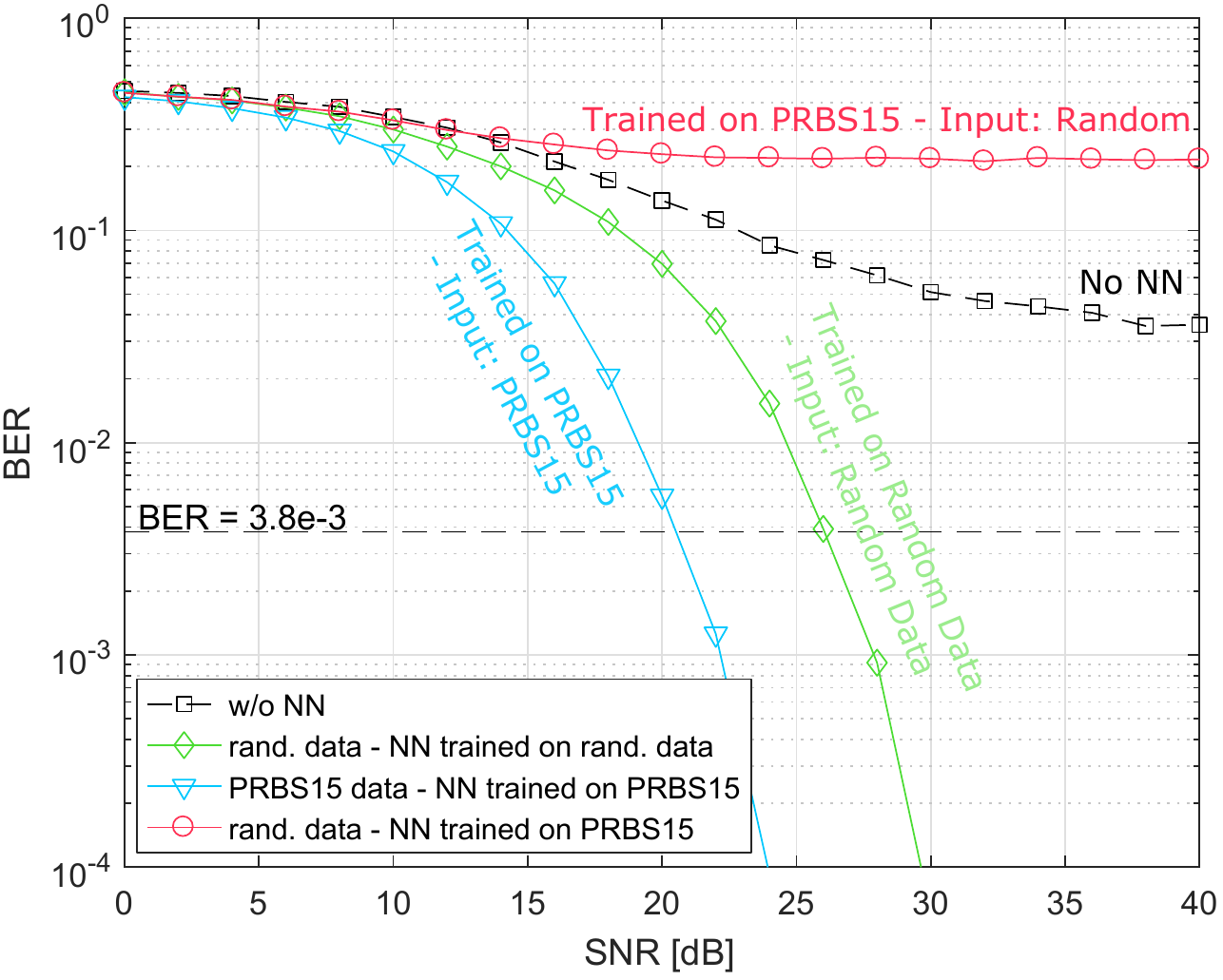}
\caption{BER vs SNR for the link in Fig.~\ref{Fig_simulationSetup} for: no neural network (black squares), the neural network trained on random data and evaluated on a different instance of random data (green diamonds), the neural network trained on PRBS15 data and evaluated on PRBS15 data (blue triangles) and the neural network trained on PRBS15 data and evaluated on random data (red circles).}
\vspace{-10pt}
\label{Fig_sweepSNR_PAM4_10kmDisp}
\end{figure}

\vspace{0pt}
\section{Application Example}
To demonstrate how this overestimation can manifest in real applications, we perform simulations for the intensity modulated direct detection (IMDD) system depicted in Fig.~\ref{Fig_simulationSetup}. We simulate 32 Gbaud PAM4 signals using raised cosine pulses with roll-off 0.95. The link is modeled with dispersion equivalent of 10~km standard single mode fiber. AWGN noise is added after square-law detection. 

The NN, the same structure as in Fig.~\ref{Fig_NN_simple}, is trained with 129 input samples (at 2 samples/symbol) using either 500000 symbols of either random data or symbols constructed from two streams of shifted PRBS15 data. 

In Fig.~\ref{Fig_sweepSNR_PAM4_10kmDisp}, the BER for the systems in Fig.~\ref{Fig_NN_simple} is plotted as a function of SNR. As seen, without the NN the system experiences an error floor at around BER = $4\times10^{-2}$. Applying the NN that has been trained on random data on a sequence of a new instance of random data, the BER can be pushed below the commonly assumed HD-FEC threshold of BER = $3.8\times10^{-3}$ when the SNR is larger than 22~dB. 

If we instead train the NN on PRBS15 data and evaluate it using PRBS15 data, we severely overestimate the gain. For this particular NN, the required SNR at a BER of $3.8\times10^{-3}$ is 5.5~dB lower compared to the case with random data. However, as discussed in the previous section, the gains comes from predicting the pattern that is used. In coherence with previous sections, when the NN trained with PRBS15 is fed with a random pattern, the performance is bad for any given SNR.

\vspace{-5pt}
\section{Discussion}
From the results presented in this paper it is obvious that great care has to be taken when evaluating NNs, especially in experiments where the pattern length is limited and where pseudo random sequences or repeated instances of random pattern are typically applied. When presenting experimental results where NNs have been applied, the following should be declared:
\begin{itemize}  
\item The pattern type that is used.
\item The length of the pattern and if it is repeated.
\item The size of the training and evaluation set.
\item If a different pattern is used for training and evaluation.
\end{itemize}

Without such information, it is impossible to ensure that the gain can be achieved with independent  data and that the gain is not a result of overfitting and/or pattern prediction. To avoid overfitting,   independent random statistics should be ensured for the training and evaluation sets.

\vspace{-0pt}
\section{Conclusion}
We have shown the dangers of overestimating the performance gain when applying NNs in experiments where PRBS sequences or repeated short random sequences are typically used. When investigating NNs in experimental work, such as for fiber optical communication, there is a need to clearly specify the strategy for training and testing, including which patterns and the length of these that were used. Without such information, it is impossible to judge if the proposed scheme are evaluated fairly and to judge if parts or all of the gain comes from predicting the pattern that was used in the study.


\ifCLASSOPTIONcaptionsoff
  \newpage
\fi



%
\vspace{-0pt}
\bibliographystyle{IEEEtran}
\bibliography{IEEEabrv,referencesNew}

%







\end{document}